\documentclass[aps,prl,twocolumn,groupedaddress,floatfix,showpacs]{revtex4}

\usepackage{graphicx}

\begin{document}


\title{Self-Calibrated Cluster Counts as a Probe of Primordial 
Non-Gaussianity}

\author{Masamune Oguri}
\email{oguri@slac.stanford.edu}
\affiliation{
Kavli Institute for Particle Astrophysics and Cosmology, Stanford
University, 2575 Sand Hill Rd. M/S 29, Menlo Park, CA 94025.
}

\date{\today}

\begin{abstract}
We show that the ability to probe primordial non-Gaussianity with
cluster counts is drastically improved by adding the excess variance
of counts which contains information on the clustering. The
conflicting dependences of changing the mass threshold and including
primordial non-Gaussianity on the mass function and biasing indicate 
that the self-calibrated cluster counts well break the degeneracy
between primordial non-Gaussianity and the observable-mass relation.
Based on the Fisher matrix analysis, we show that the count variance
improves constraints on $f_{\rm NL}$ by more than an order of
magnitude. It exhibits little degeneracy with dark energy equation of
state. We forecast that upcoming Hyper Suprime-cam cluster surveys and
Dark Energy Survey will constrain primordial non-Gaussianity at the
level $\sigma(f_{\rm NL})\sim 8$, which is competitive with
forecasted constraints from next-generation cosmic microwave
background experiments. 
\end{abstract}


\maketitle

The measurement of departures from Gaussianity of the initial
perturbations provides a unique opportunity to probe the early
universe \cite{Bartolo:2004if}. While the standard single field
slow-roll inflation models predict primordial perturbations very 
close to Gaussian, some models such as multi-field models and the
curvaton model can produce the level of non-Gaussianity high enough to
be detected in ongoing or future surveys.  Thus specific forms of
primordial non-Gaussianity contain valuable information on how the
initial density fluctuations are generated.  

Observationally, primordial non-Gaussianity has mainly been studied 
using the temperature fluctuation of the Cosmic Microwave Background
(CMB). Recently it has attracted considerable attention given a
possible detection of non-Gaussianity by Yadav and Wandelt
\cite{Yadav:2007yy}. However, the detection of non-Gaussianity in the
CMB is somewhat controversial in the sense that independent analyses
yield slightly different results \cite{cmb}, suggesting the importance
of other observational probes independent of the CMB. Another powerful
probe of primordial non-Gaussianity is provided by the large-scale
structure of the universe. In particular,  non-Gaussianity induces a 
scale-dependent halo bias
\cite{Dalal:2007cu,Matarrese:2008nc,Slosar:2008hx,Afshordi:2008ru,Desjacques:2008vf}, 
and thus by studying large-scale distributions of astronomical objects
one can obtain tight constraints that are competitive with the CMB.
Constraints from the large-scale structure are also important given
that non-Gaussianity can be scale-dependent such that deviations from
Gaussian are larger at smaller scales \cite{LoVerde:2007ri}. 

Primordial non-Gaussianity is also sensitive to the abundance of
massive clusters and its redshift evolution
\cite{Dalal:2007cu,Sefusatti:2006eu}. An advantage of using massive
clusters is its reasonable one-to-one correspondence with dark halos,
which suggests that halo assembly bias (e.g., \cite{Slosar:2008hx})
is less important. A challenge here is how to calibrate cluster
masses; since the cluster mass is not directly observable, one has to
resort to well-calibrated correlations between cluster masses and
observable quantities such as luminosities, temperatures, and the
numbers of member galaxies in order to infer cluster masses. The
observable-mass relations always involve uncertainties, suggesting
that the change of cluster abundances by primordial non-Gaussianity
may be compensated by modifying the relation between observables and
masses. Therefore constraints from cluster counts rely on how well we
can calibrate such observable-mass relations.  

In this {\it Letter}, we point out that clustering information breaks
the degeneracy and allows us to determine primordial non-Gaussianity
surprisingly well with cluster counts. This is because the clustering
bias for massive clusters is quite sensitive to both cluster masses and
primordial non-Gaussianity, and more importantly, because the cluster
abundance and biasing show conflicting dependences on these. Such
self-calibrated cluster count technique has been discussed extensively
in the context of accurate dark energy probes
\cite{Majumdar:2003mw,Lima:2004wn,Lima:2005tt,Lima:2007kx}, but its
use for primordial non-Gaussianity has not been explored. 

Here we quantify non-Gaussianity of the local form using the standard 
parametrization,  
$\Phi=\phi+f_{\rm NL}\left(\phi^2-\langle\phi^2\rangle\right)$, 
where $\Phi$ is the curvature perturbation and $\phi$ is an auxiliary
random-Gaussian field. The parameter $f_{\rm NL}>0$ ($<0$) indicates
that the initial density field is positively (negatively) skewed. 
The current level of constraints on primordial non-Gaussianity is
$|f_{\rm NL}| \lesssim\mathcal{O}(100)$ \cite{Yadav:2007yy,cmb,Slosar:2008hx}. 
We adopt a non-Gaussian correction factor of the cluster mass function
based  on the Edgeworth expansion \cite{LoVerde:2007ri}:
\begin{equation}
\frac{dn/dM}{dn_{\rm G}/dM}=1+\frac{\sigma S_3}{6}
(\nu^3-3\nu)-\frac{1}{6}\frac{d(\sigma S_3)}{d\ln\nu}\left(
\nu-\frac{1}{\nu}\right),
\end{equation}
where $\nu=\delta_c/\sigma$, $\delta_c \approx 1.68$ is the critical
linear overdensity, $\sigma=\sigma(M,z)$ is the linear fluctuation on
the mass scale of $M$ which we compute using the transfer function
$T(k)$ presented by Eisenstein and Hu \cite{Eisenstein:1997ik}
ignoring the baryon wiggle. We adopt models of Warren {\it et al.}
\cite{Warren:2005ey} for the mass function in the Gaussian case,
$dn_{\rm G}/dM$. The skewness $S_3$ is related to $f_{\rm NL}$ as
\cite{Desjacques:2008vf}
\begin{eqnarray}
\sigma S_3&=&\frac{f_{\rm NL}}{\sigma^3}\int_0^\infty\frac{dk_1}{k_1}\alpha(k_1)W(M,
k_1)\Delta_\phi^2(k_1)\nonumber\\
&&\times\int_0^\infty\frac{dk_2}{k_2}\alpha(k_2)W(M,
k_2)\Delta_\phi^2(k_2)\nonumber\\
&&\times\int_{-1}^1d\mu\,\alpha(k)W(M, k)\left[1+\frac{P_\phi(k)}{P_\phi(k_1)}+\frac{P_\phi(k)}{P_\phi(k_2)}\right],
\end{eqnarray}
where $\Delta^2_\phi=k^3P_\phi(k)/2\pi^2$ is the power spectrum of the
curvature perturbation, $\alpha=[2D(z)T(k)/3\Omega_M](ck/H_0)^2$,
$D(z)$ is the linear growth rate normalized to $(1+z)^{-1}$ in the
matter-dominant era, and $k^2=k_1^2+k_2^2+2\mu k_1k_2$. For the window
function $W(M, k)$ we adopt the real space top-hat filter. In practice
we use the following fitting formula for $\sigma S_3$:  
\begin{eqnarray}
\sigma S_3&\approx&(8.66\times 10^{-5})f_{\rm
  NL}\frac{\Omega_M}{D(0)}\Gamma^{-1.4}\sigma_8\nonumber\\
&&\times m_{10}^{-0.0272-0.11(n_s-0.96)-0.0008\log m_{10}},
\end{eqnarray}
with 
$m_{10}=[M/(10^{10}h^{-1}M_\odot)]\Gamma^3(\Omega_Mh^2)^{-1}$
and $\Gamma=\Omega_Mh\exp[-\Omega_b(1+\sqrt{2h}/\Omega_M)]$
is so-called the shape parameter \cite{Sugiyama:1994ed}. This fitting
formula should be accurate at a few percent level in the mass scale
range $10^{7}h^{-1}M_\odot\lesssim M\lesssim 10^{18}h^{-1}M_\odot$. 

\begin{figure}[!t]
\includegraphics[width=0.48\textwidth]{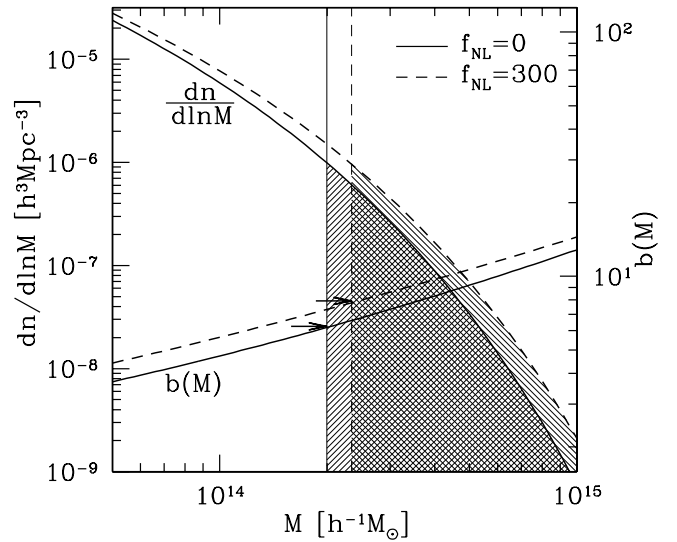}
\caption{The illustration of self-calibrating cluster counts to probe
  primordial non-Gaussianity $f_{\rm NL}$. Here the redshift is
  $z=1$. In cluster surveys, we basically obtain the number of
  clusters above some mass threshold $M_{\rm th}$. The plot indicates
  that the increase of the mass function $dn/dM$ due to positive
  $f_{\rm NL}$ can be compensated by increasing the $M_{\rm
  th}$. However, while the two models predict the same number of
  clusters, the corresponding halo bias $b(M)$ (here we adopted
  $k=0.02h{\rm Mpc^{-1}}$) are quite different because both raising
  $M_{\rm th}$ and $f_{\rm NL}$ increase $b(M)$ (see arrows). 
  Although in this plot the scatter in the observable-mass relation is
  ignored for simplicity, it will be included in the Fisher matrix
  analysis. 
} 
\label{fig:nb}
\end{figure}

The non-Gaussian correction of the halo bias is computed as 
\cite{Desjacques:2008vf}
\begin{equation}
\Delta b(M,z,k)=\frac{2f_{\rm NL}\delta_c}{\alpha}\left(b_{\rm
    G}-1\right)-\frac{\nu}{\delta_c}\frac{d}{d\nu}\left(\frac{dn/dM}{dn_{\rm
      G}/dM}\right), 
\end{equation}
The halo bias in the Gaussian case, $b_{\rm G}$, is assumed to be the
form presented by Sheth and Tormen \cite{Sheth:1999mn}.

Fig.~\ref{fig:nb} illustrates the reason why the clustering
information is so important. As shown in the Figure, including
positive $f_{\rm NL}$ increases the number of clusters above some mass 
threshold $M_{\rm th}$. This increment can be compensated by raising
$M_{\rm th}$. However, these two models with the same numbers of
clusters result in quite different halo biases because {\it both}
raising $M_{\rm th}$ and $f_{\rm NL}$  increase the biasing. Thus by
including clustering information we can strongly break the degeneracy
between $M_{\rm th}$ and  $f_{\rm NL}$, and can obtain tight
constraints on $f_{\rm NL}$.  

We now forecast constraints on $f_{\rm NL}$ from future cluster
surveys. We include the clustering information using a count-in-cell
analysis. Specifically we approximate the Fisher matrix as
\cite{Lima:2004wn,Lima:2005tt}  
\begin{equation}
F_{\alpha\beta}=\mathbf{m}^T_{,\alpha}\mathbf{C}^{-1}\mathbf{m}_{,\beta}+\frac{1}{2}{\rm
  Tr}\left[\mathbf{C}^{-1}\mathbf{S}_{,\alpha}\mathbf{C}^{-1}\mathbf{S}_{,\beta}\right]+\frac{\delta_{\alpha\beta}}{\sigma_{\rm p}^2(\alpha)},
\label{eq:fish}
\end{equation}
where $\sigma_{\rm p}$ represents the prior information on each
parameter, and the covariance matrix is given by
$\mathbf{C}\equiv\mathbf{S}+{\rm diag}(\mathbf{m})$. The number count
$\mathbf{m}$ and its variance $\mathbf{S}$ are computed as
\begin{equation}
m_i=V_i\int_i \frac{dM_{\rm obs}}{M_{\rm obs}}\int dM\frac{dn}{dM}p(M_{\rm obs}|M),
\end{equation}
\begin{equation}
S_{ij}=\frac{1}{V_iV_j}\int
\frac{d^3k}{(2\pi)^3}W^*_i(\mathbf{k})W_j(\mathbf{k})P(k) b_ib_j,
\end{equation}
\begin{equation}
b_i=V_i\int_i \frac{dM_{\rm obs}}{M_{\rm obs}}\int dM\frac{dn}{dM}b(M,k)p(M_{\rm obs}|M),
\end{equation}
where the subscript $i$ run over redshift, mass, and angular bins. 
The power spectrum is described by $P(k)$, and the $k$-space window
function by $W_i(\mathbf{k})$. Since the off-diagonal elements of
$\mathbf{S}$ are small in our case, here we consider only the diagonal
elements. The function $p(M_{\rm obs}|M)$ models the accuracy of the
cluster mass determination from observables. Following
\cite{Lima:2005tt}, we assume the log-normal distribution for
$p(M_{\rm obs}|M)$, with the median of $\ln M +\ln M_{\rm bias}$ and
the scatter of $\sigma_{\ln M}$, and regard $\sigma_{\ln M}$ and $\ln
M_{\rm bias}$ (which corresponds to $M_{\rm th}$ in Fig.~\ref{fig:nb})
as nuisance parameters. Note that the first term of the 
Fisher matrix (Eqn. \ref{eq:fish}) represents the information from
number counts, whereas the second term the information from the
variance of the counts which contain clustering (biasing)
information. Using the Fisher matrix, one can estimate a marginalized
error on each parameter as 
$\sigma(\alpha)=\sqrt{(\mathbf{F}^{-1})_{\alpha\alpha}}$. 

We calculate the Fisher matrix in 10-dimensional parameter space; 6
standard cosmological parameters including dark energy equation of
state (the matter density $\Omega_M h^2$, the baryon density $\Omega_b
h^2$, the power spectrum tilt $n_s$, the normalization of the power
spectrum $\delta_\zeta$ \cite{Hu:2003pt}, the dark energy density
$\Omega_{\rm  DE}$, and dark energy equation of state $w$), 1 parameter
representing primordial non-Gaussianity ($f_{\rm NL}$), and 3
parameters from the observable-mass relation, $\sigma_{\ln M}$ and
$\ln M_{\rm bias}=\ln M_{\rm bias,0}+\gamma\ln(1+z)$. The
Five-Year Wilkinson Microwave Anisotropy 
Probe (WMAP5) result for $\Lambda$CDM 
\cite{Dunkley:2008ie}, ($\Omega_M h^2$, $\Omega_b h^2$, $n_s$,
$\delta_\zeta$, $\Omega_{\rm DE}$, $w$)=(0.133, 0.0227, 0.963,
$4.61\times 10^{-5}$, 0.742, $-1$), is adopted as our fiducial
cosmological model. We add 
conservative priors to the first 4 parameters, $\sigma_{\rm p}(\Omega_M
h^2)=0.006$, $\sigma_{\rm p}(\Omega_b h^2)=0.0006$, 
$\sigma_{\rm p}(n_s)=0.015$, and $\sigma_{\rm p}(\delta_\zeta)=10^{-6}$; 
these are the level of accuracies which has already been achieved by
WMAP5. In addition, our fiducial model has $f_{\rm NL}=0$,  $\sigma_{\ln
  M}=0.25$, $\ln M_{\rm bias,0}=0$, and $\gamma=0$. 

For illustrative purposes, we consider the following three upcoming 
surveys; Hyper Suprime-Cam on Subaru telescope (HSC;
since the design of the HSC cluster survey is still tentative, we
consider both 1000~deg$^2$ and 2000~deg$^2$), 
Dark Energy Survey (DES; 5000~deg$^2$) \cite{des}, and Large Synoptic
Survey Telescope (LSST; 20000~deg$^2$) \cite{lsst}. We adopt a
simplified assumption that these optical surveys will find clusters
with $M_{\rm   obs}>10^{13.7}h^{-1}M_\odot$ out to $z_{\rm max}=1.4$,
$1.0$, and $1.7$, respectively. For the count-in-cell analysis, we use
the cell size of 20~deg$^2$ and the redshift interval $\Delta
z=0.1$. Three mass bins with spacing of $\Delta\log M_{\rm obs}=0.5$
are also adopted.   

\begin{figure}
\includegraphics[width=0.42\textwidth]{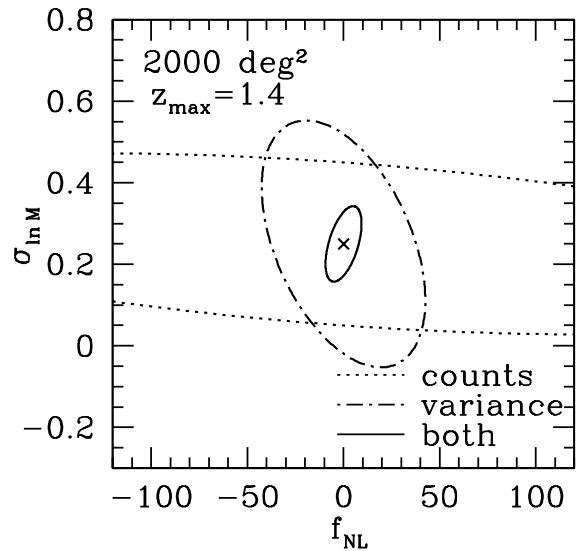}
\caption{Expected marginalized constraints in the 
  $f_{\rm NL}$-$\sigma_{\ln M}$ plane from the upcoming 2000~deg$^2$
  HSC cluster survey. WMAP5 cosmology is assumed as a fiducial
  model. Contours indicate the 68\% confidence regions from the number
  counts ({\it dotted}), the variance of the counts which contains the
  clustering information ({\it dash-dotted}), and the combination of
  the number counts and variance ({\it solid}).}
\label{fig:cons}
\end{figure}

In Fig.~\ref{fig:cons}, we show marginalized constraints on $f_{\rm
 NL}$ and the correlations with the parameters $\sigma_{\ln M}$,
expected for the 2000~deg$^2$ HSC cluster survey.  As expected,
constraints are drastically improved by combining the number counts
with the variance which includes the clustering information. 
This is partly because constraints from number counts and variance
show different degeneracy directions, suggesting that {\it both} the
number counts and clustering are essential for accurate determinations
of $f_{\rm NL}$.

\begin{table*}[!t]
\begin{tabular}{@{}c|ccc|ccc|ccc}
  \hline\hline
 & \multicolumn{3}{|c|}{Counts} & \multicolumn{3}{|c|}{Variance} & \multicolumn{3}{|c}{Counts + Variance}\\ \cline{2-10}
Survey & $\sigma(\Omega_{\rm DE})$ & $\sigma(w)$ & $\sigma(f_{\rm
 NL})$ &  $\sigma(\Omega_{\rm DE})$ & $\sigma(w)$ & $\sigma(f_{\rm
 NL})$ & $\sigma(\Omega_{\rm DE})$ & $\sigma(w)$ & $\sigma(f_{\rm NL})$ \\ \hline
HSC1 & 0.030 & 0.151 & 240.4 & 0.012 & 0.189 & 37.8 & 0.010 & 0.103 & 8.1 \\
HSC2 & 0.023 & 0.108 & 188.2 & 0.011 & 0.142 & 28.0 & 0.009 & 0.074 & 6.2 \\
DES  & 0.032 & 0.081 & 210.6 & 0.011 & 0.102 & 35.3 & 0.009 & 0.055 & 7.9 \\
LSST & 0.009 & 0.037 & 106.1 & 0.006 & 0.051 &  6.9 & 0.005 & 0.024 & 1.9 \\ \hline
\end{tabular}
\caption{Marginalized constraints on cosmological parameters estimated
from the Fisher matrix analysis using the number counts and/or the
variance of counts (clustering).
WMAP5 cosmology is assumed as a fiducial model. Constraints in four
future survey parameters, HSC1 (1000~deg$^2$, $z_{\rm max}=1.4$), HSC2
(2000~deg$^2$, $z_{\rm max}=1.4$), DES (5000~deg$^2$, $z_{\rm
  max}=1.0$), and LSST (20000~deg$^2$, $z_{\rm max}=1.7$), are presented.} 
\label{tab:cons}
\end{table*}

\begin{figure}
\includegraphics[width=0.45\textwidth]{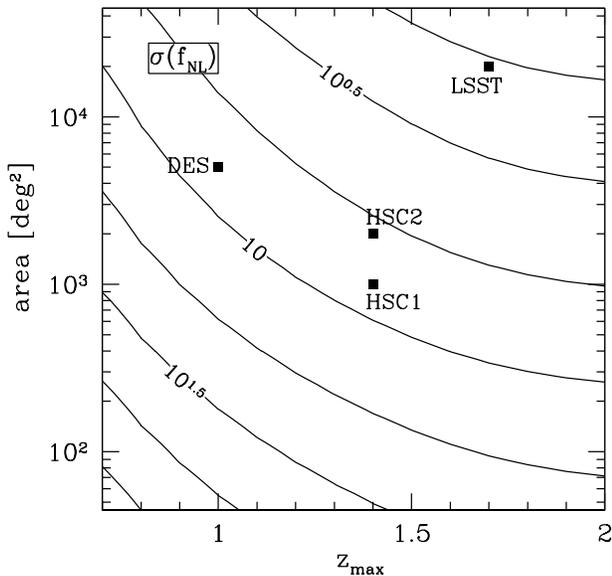}
\caption{Expected marginalized constraints on $f_{\rm NL}$ as a
  function of two survey parameters, the survey area and the maximum
  redshift $z_{\rm max}$. Contours are drawn per 0.25~dex. 
  Four survey parameters considered in this
  paper are indicated by filled squares.}  
\label{fig:cont_fnl}
\end{figure}

Table~\ref{tab:cons} summarizes forecasted constraints on various
cosmological parameters. For all the upcoming surveys, the count
variance drastically enhances the ability to probe primordial
non-Gaussianity, by more than an order of magnitude improvement in
$f_{\rm NL}$ compared with the number counts alone. Predicted
marginalized errors of $\sigma (f_{\rm NL})\sim 8$ for HSC and DES
and $\sim 2$ for LSST are competitive with constraints from
next-generation CMB experiments (e.g., \cite{Serra:2008wc}) and
galaxy power spectrum measurements (e.g.,
\cite{Dalal:2007cu,Carbone:2008iz}). The variance helps to
regulate the observable-mass relation, improving an accuracy of $\ln
M_{\rm bias,0}$ and $\sigma_{\ln M}$ measurements by a factor of two
or more. Measurements of dark energy equation of state are improved as
well, which is consistent with earlier work. We find that $w$ and
$f_{\rm NL}$ are not correlated very much, indicating that we can well
determine these two parameters simultaneously using self-calibrated
cluster counts.  Fig.~\ref{fig:cont_fnl} shows contours of $\sigma
(f_{\rm NL})$ as a function of the survey area and the maximum redshift. 
The expected constraints on $f_{\rm NL}$ is a steep function of the
maximum redshift even at $z>1$, which indicate the importance of the
deep surveys to detect clusters out to $z\gtrsim 1$. 

We have shown that adding clustering information from the count
variance drastically improves measurements of primordial
non-Gaussianity with cluster counts. Although the calibration of
cluster masses limits the use of cluster counts as a cosmological
probe, the self-calibration technique allows us to determine both the
observable-mass relation and $f_{\rm NL}$ simultaneously. The
significant effect of the count variance comes from the conflicting
dependences of the mass threshold and $f_{\rm NL}$ on the cluster mass
function and biasing parameter (Fig.~\ref{fig:nb}). Allowing
dark energy equation of state to vary does not degrade $f_{\rm NL}$
measurements very much. Resulting forecasted constraints on
$f_{\rm NL}$,  $\sigma (f_{\rm NL})\sim 8$ for HSC and DES and $\sim
2$ for LSST, suggest that cluster counts can become a competitive
probe compared to the CMB or the large-scale galaxy power spectrum. 

We have here made a number of simplified assumptions. For instance, it
is important to check how the possible redshift evolution of the
observable-mass relation affects our results
\cite{Lima:2004wn,Lima:2005tt}. The impact of other systematics, such
as cluster photometric redshifts \cite{Lima:2007kx} and the effect of
halo assembly bias \cite{Wu:2008ij}, should be addressed. On the other
hand, we used only the count variance of each cell as the clustering
information. Since the effect of primordial non-Gaussianity is more
significant at larger scales, including the count covariance (or
including the full clustering information with the power spectrum
\cite{Majumdar:2003mw}) may improve the constraints further. We leave
such more comprehensive treatments for future work.  

\begin{acknowledgments}
This work was supported by Department of Energy contract DE-AC02-76SF00515.
\end{acknowledgments}

\end{document}